\documentclass[12pt]{article}

\newcommand{\beq}{\begin{equation}}
\newcommand{\eeq}{\end{equation}}
\newcommand{\beqa}{\begin{eqnarray}}
\newcommand{\eeqa}{\end{eqnarray}}
\newcommand{\beqar}{\begin{eqnarray*}}
\newcommand{\eeqar}{\end{eqnarray*}}

\newcommand{\al}{\alpha}
\newcommand{\be}{\beta}

\def\spa          {\ \ \ }
\def\non          {\nonumber}
\def\ha           {\mbox{$\frac{1}{2}$}}

\def\spa          {\ \ \ }
\def\mand         {\spa\mbox{and}\spa}

\def\Tr           {\mbox{\rm Tr}\,}
\def\STr          {\mbox{\rm STr}\,}

\def\cd           {{\cdot}}
\def\ran          {\rangle}
\def\lan          {\langle}

\def\fsH    {H\!\!\!\!/\,}
\def\fsk    {k\!\!\!\!/\,}

\newcommand{\del}{\delta}

\newcommand{\eps}{\epsilon}
\newcommand{\ga}{\gamma}

\newcommand{\inn}{\!\cdot\!}

\newcommand{\lam}{\lambda}

\newcommand{\z}{\zeta}

\newcommand{\labell}[1]{\label{#1}} 
\newcommand{\reef}[1]{(\ref{#1})}
\newcommand\prt{\partial}

\newcommand\cD{{\cal D}}

\newcommand\bD{\bar{D}}

\def\sst#1{{\scriptscriptstyle #1}}
\def\0{{\sst{(0)}}}
\def\1{{\sst{(1)}}}
\def\2{{\sst{(2)}}}
\def\3{{\sst{(3)}}}
\def\4{{\sst{(4)}}}
\def\5{{\sst{(5)}}}
\def\6{{\sst{(6)}}}
\def\7{{\sst{(7)}}}
\def\8{{\sst{(8)}}}

\begin{document}
\baselineskip 18pt%
\begin{titlepage}

\center{ {\bf \Large   All Order $\alpha'$ Corrections to Fermionic Couplings \\
of D-brane-Anti-D-brane Effective Actions

}}\vspace*{3mm} \centerline{{\Large {\bf  }}}
\begin{center}
{ Ehsan Hatefi$^{a,b}$ }

\vspace*{0.6cm}{

Scuola Normale Superiore and INFN,\\
 Piazza dei Cavalieri, 7, 56126 Pisa, Italy$^{a}$ \\
\quad\\

Mathematical Institute, Faculty of Mathematics,\\
Charles University, P-18675, CR$^{b,}$\footnote{ ehsan.hatefi@sns.it, e.hatefi@qmul.ac.uk, ehsanhatefi@gmail.com}  }

\end{center}
\begin{center}{\bf Abstract}\end{center}
\begin{quote}

We first point out to  all order $\alpha'$ corrections to lower order fermionic couplings  and their effective actions of type II super string theory. In order to reveal all symmetries of the particular D-brane Anti-D-brane system, and to construct not only all the singularity structures but also all order  $\alpha'$ corrections of fermionic couplings to tachyons, we employ all the Conformal Field Theory (CFT) methods to a six point function. Basically,  we construct all the correlation functions of four spin operators and two world sheet fermion fields. The S-matrix of two fermion fields, two real tachyons in the presence of a closed string Ramond-Ramond (RR) field in type II will be explored. Applying all the symmetries of the amplitude, a proper expansion for D-brane Anti D-brane is also found out. Having carried out
all the  Effective Field Theory (EFT) techniques, one would discover all order singularity structures of string theory. The entire algebraic computations of the amplitude of  $<V_{C^{-1}(z,\bar z)}  V_{\bar\psi^{-1/2}(x_1)}  V_{\psi^{-1/2}(x_2)} V_{T^{0}(x_3)} V_{T^{0}(x_4)}>$ have also been derived. Various new string couplings are revealed as well. Eventually, making comparison the S-matrix with all EFT parts,  all order $\alpha'$ higher derivative corrections to two fermions-two tachyons in the world volume of  D-brane Anti D-brane of type II are discovered accordingly.

 \end{quote}
\end{titlepage}

\section{Introduction}

In order to extract new insight about special features of superstring theory and also to talk about 
 the structures as well as exact coefficients of the string theory corrections, one needs to properly deal with  S-matrix formalism in both Quantum Field Theory and String Theory.  By exploring the amplitudes of  non-supersymmetric cases, one might be able to actually realise how supersymmetry gets broken. More crucially, one explores new effective couplings in both Effective Field Theory (EFT) and String theory, and in particular in time dependent backgrounds as well. To do so, we aim to function just with scattering amplitude methods in the world volume of D-brane Anti-D-brane system of type II. To understand some important  properties of string couplings we refer the interested reader to have a look at the following original works in various different context \cite{Gutperle:2002ai,Lambert:2003zr,Sen:2004nf}. 

\vskip.2in

To illustrate open strings and their crucial roles in revealing many potential different facets of strings, we highlight the most original references in \cite{orientifold}.


\vskip.2in

D-brane Anti-D-brane Effective Actions of string theory have potential applications for various holographic models such as Sakai- Sugimoto model \cite{Sakai:2004cn}, also do have application to the  physics of symmetry breaking for different QCD models \cite{Casero:2007ae}. Indeed one starts to embed some flavour branes in parallel to branes and consequently Anti branes are regarded as some backgrounds that might be dual to colour confined systems. Notice that this system is also assumed to be a probe if and only if the number of flavours are much less that colour ones $N_f<< N_c$. Open string tachyons do play special roles in addressing instability of different systems, therefore it is needed to verify the structure and coefficients of these modes in EFT as well as their all higher order interactions in type II string theory. Employing Conformal Field Theory (CFT) methods, would make it possible to clarify in detail their effective actions with great care as well.
\vskip.2in

A. Sen and other people have already found the tachyonic effective actions at lowest order \cite{Sen:1999md,Bergshoeff:2000dq} and those effective actions would provide some information about properties of the system such as  decaying  the non-BPS branes \cite{Sen:2002in} and etc. Having considered \cite{Sen:2004nf}, one would come to know how to embed massless strings as well as tachyons in an EFT part. We started addressing all order effective actions of non-BPS branes and in particular D-brane-Anti-D-brane system with great care in \cite{Hatefi:2017ags} where tachyon condensation for this system was also understood by A. Sen in  \cite{Sen:1998sm}.

\vskip.2in

Two real tachyon modes do appear if and only if one goes through the places where the distance between brane and Anti-brane picks up the particular value of less than string 's length scale where one replaces them out in an EFT and evidently starts to take into consideration their dynamics. Their new couplings can just be explored by direct scattering amplitude methods not by any duality transformation whatsoever. To employ their dynamics one could consider one of the latest works done by the late Joe Polchinski and his collaborators \cite{Michel:2014lva} where they treated  D-brane effective action in such a way that standard EFT parts make sense. 

\vskip.2in

Note that the known Four-Dimensional Superstrings were discovered in \cite{Antoniadis:1986rn}, then the gauged supergravity vacua in string theory were realised in \cite{Antoniadis:1989mn} and later on the BPS and non BPS branes of various string 
theories with or without broken supersymmetry were constructed in \cite{Dudas:2001wd}. To deal with the stability issues with broken supersymmetry, we refer the interested reader to some of the devoted papers as appeared in \cite{Dudas:2004nd,Basile:2018irz} and also to the very interesting and recent comprehensive review  appeared in \cite{Mourad:2017rrl}.

\vskip.2in

To talk about the applications of non-BPS branes, one could mention many references, however, we just decided to confide producing branes \cite{Bergman:1998xv}, describing some inflationary scenarios in string theory, and refer to inflation in string theory in the language explained in  KKLT \cite{Dvali:1998pa}.

\vskip.2in

Using scattering amplitude techniques, not only does one explore new effective couplings but also can fix exact coefficients of all higher derivative couplings, more crucially for once, one would be able to fix all the ambiguities that appear at $\alpha'$ corrections of both type II string theory. \footnote{Look at \cite {Bjerrum-Bohr:2014qwa} to see some of the higher order corrections and to distinguish all order BPS corrections we refer the interested reader in observing \cite{Hatefi:2012zh,Hatefi:2012ve}. To  notify all order $\alpha'$  corrections to BPS branes one refers to a universal conjecture that is demonstrated in detail in \cite{Hatefi:2012rx}, which can also be applied to non-BPS branes where some further ingredients are needed.}

 One might be able to even talk about various thermodynamical aspects of this particular system, which can be stabilised at finite temperature and may have potential applications to black holes \cite{AlvarezGaume:2011rk} as well as various applications to famous  AdS/CFT conjecture or even to the known M-theory approach \cite{Hatefi:2012bp}. It is also interesting to remark that, D-brane Anti D-brane system has also something to do with stability of KKLT or a particular string compactification in  Large Volume String Scenario \cite{Polchinski:2015bea}. Thanks to Pioneer works, the famous correspondence between D-branes and their sources being RR fields has been largely discussed in many reviews but we just point out to the original one that was made by the master of D-branes \cite{Polchinski:1995mt} and also their bound states  are further argued in \cite{Witten:1995im}. 

\vskip.2in

Since we are all aware of the fact that duality does not hold for non-BPS branes, one needs to work with scattering amplitude methods to be able to fix precise $\alpha'$ corrections and that can be carried out if and only if one employs the direct CFT methods, where various favours have been set out in \cite{Hatefi:2012zh}. 
\vskip.2in

Notice that we have explained all three known EFT ways of exploring couplings that are known to be thought of either Wess-Zumino (WZ) actions for both BPS and non-BPS DBI effective actions \cite{Hatefi:2010ik} or  Myers terms \cite{Myers:1999ps} or Pull back of branes where Taylor expansion of the standard transverse scalar fields is also taken.


\vskip.2in

The paper is organised as follows. As a warm-up, we first point out to lowest order D-brane-Anti-D-brane effective actions in the presence of fermion fields and also talk about various Bianchi identities and other potentially different effective couplings that are found in both bulk and world volume directions of this important system.  Note that the authors in \cite{Kennedy:1999nn} were the first scientists whom explored some important couplings at lowest orders. By emerging CFT methods as well as scattering amplitude formalism, we could precisely deal with their higher orders and showed how to properly explore not only various new couplings but also their all order  $\alpha'$ higher derivative corrections in both type II string theory.

\vskip.1in

To deal with symmetries, by making use of CFT methods \cite{Friedan:1985ge}
 we first reveal the S-matrix elements of a six point function of two fermion fields (that are massless and are located either on world volume of D-brane or Anti D-brane), two real tachyons in the presence of a closed string RR field in type II.  We also try to apply the so called selection rules \cite{Hatefi:2013yxa} and simultaneously keep track of all the symmetries of the amplitude so that a proper expansion for D-brane Anti D-brane can be figured out. Having done all the proper EFT techniques, one would discover all order  singularity structures of this system. The entire algebraic computations of the amplitude  of  $<V_{C^{-1}(z,\bar z)}  V_{\bar\psi^{-1/2}(x_1)}  V_{\psi^{-1/2}(x_2)} V_{T^{0}(x_3)} V_{T^{0}(x_4)}>$ are performed.  Consequently, all the integrals in the soft limit \footnote{The soft limit for mixed amplitudes including RR and bosons is shown to be $4k_2.p\rightarrow 1$ in \cite{Hatefi:2018qlm}, while here due to the presence of fermion fields and having different correlators, it changes to $4k_2.p\rightarrow 2$ where all massless poles are also derived  in an EFT and the method of the derivation of integrals is recently explained in \cite{Antonelli:2019pzg}.} are obtained.  Finally all order $\alpha'$ higher derivative corrections to two fermion fields and two tachyons in the world volume of D-brane Anti D-brane system in type II string theory are discovered.

\vskip.1in

The first step is to properly explore all the correlation functions of four spin operators with two world sheet fermion fields $<V_{C^{-1}(z,\bar z)}  V_{\bar\psi^{-1/2}(x_1)}  V_{\psi^{-1/2}(x_2)} V_{T^{0}(x_3)} V_{T^{0}(x_4)}>$ in both type II string theory, where along the way various technical issues will be addressed as well. 
 In order to expand the final form of the S-matrix we are going to use the so called selection rules \cite{Hatefi:2013yxa}, all lower point functions of D-brane Anti D-brane system \cite{Hatefi:2016yhb}, standard  EFT couplings, the particular soft limit as well as the extra symmetries that are elaborated later on. We then would be able to reconstruct  all infinite massless and tachyon singularities in both EFT and string part as well.  Eventually one comes to know that all the higher derivative corrections of two fermions- two tachyons need to be taken into account to be able to precisely find out the singularities that are advocated by string amplitude. Due to selection rules and the internal degrees of freedom or the Chan-Paton factors, it is known that there is no interaction between two fermions and  a tachyon, hence all three point functions of a tachyon and two fermions (with both the same and different chirality)  are zero. Therefore, as it will be observed from the ultimate form of the amplitude we have no longer singularity structures in $s',v',u',r', (s'+v'+w'),(u'+r'+w')$ channels either.

 \section{Lower Order Fermionic Effective Actions }

 In order to proceed we first write down the vertex operators with their ingredients and then warm up with some  effective couplings as well as their corrections that are useful in the other sections. Vertex operators and their Chan-Paton (CP) factors for D-brane Anti-D-brane system are introduced as follows
  \beqa
 V_{T}^{(0)}(x) &=&  \alpha' ik_3\cd\psi(x) e^{\alpha' ik_3. X(x)}\lam\otimes\sigma_1,
\nonumber\\
V_{T}^{(-1)}(x) &=&e^{-\phi(x)} e^{\alpha' ik\cd X(x)}\lam\otimes\sigma_2\nonumber\\
V_{\bar\Psi}^{(-1/2)}(x_1)&=&\bar u^Ae^{-\phi(x_1)/2}S_A(x_1)\,e^{ \alpha'ik_1.X(x_1)}\lam\otimes\sigma_3 \nonumber\\
V_{\Psi}^{(-1/2)}(x_2)&=&u^Be^{-\phi(x_2)/2}S_B(x_2)\,e^{ \alpha'  ik_2.X(x_2)}\lam\otimes I \nonumber\\
V_{C}^{(-\frac{1}{2},-\frac{1}{2})}(z,\bar{z})&=&(P_{-}\fsH_{(n)}M_p)^{\alpha\beta}e^{-\phi(z)/2}
S_{\al}(z)e^{i\frac{\alpha'}{2}p\cd X(z)}e^{-\phi(\bar{z})/2} S_{\be}(\bar{z})
e^{i\frac{\alpha'}{2}p\cd D \cd X(\bar{z})} \otimes\sigma_3\nonumber
\eeqa

 \vskip.1in
 
Our notations are shown by the followings. $\mu, \nu = 0, 1, ..., 9 $ run the ten dimensional space-time, $a, b, c = 0, 1, ..., p$ and 
$ i, j = p + 1, ..., 9$ are  representing world volume and transverse directions of the brane accordingly.
  $\lambda$ shows the external CP factor in the
U(N) gauge group where $\sigma_i$ are the known Pauli Matrices. As was argued all vertices of non-BPS branes carry some degrees of freedom, because if we set the real tachyons to zero then one must be able to recover the entire Wess-Zumino effective actions that are just related to BPS branes.  Further explanations about CP factors can be read off from \cite{Hatefi:2013mwa}.

 \vskip.1in

  On-shell conditions  are
\beqa
  k_1^2=k_2^2=p^2=0, \quad  k_{3}^2=\frac{1}{2\alpha'} , \quad  \bar u^A  \fsk_{1}= \fsk_{2} u^B =0,
\nonumber\eeqa
The projection operator and also the field strength of RR  are shown as \begin{displaymath}
P_{-} =\ha (1-\ga^{11}), \quad
\fsH_{(n)} = \frac{a
_n}{n!}H_{\mu_{1}\ldots\mu_{n}}\ga^{\mu_{1}}\ldots
\ga^{\mu_{n}},
\non\end{displaymath}

where one can just point out to the spinor notation as follows:
 $ (P_{-}\fsH_{(n)})^{\al\be} =
C^{\al\del}(P_{-}\fsH_{(n)})_{\del}{}^{\be}$, where C is defined to be the charge conjugation matrix and we choose $n=2,4$,$a_n=i$  ($n=1,3,5$,$a_n=1$) for IIA (IIB) appropriately. Let us deal with  just the meromorphic parts of the field and in order to do so we apply the so called doubling trick, which means that some proper change of variable to all fields is taken into consideration as 
\begin{displaymath}
\tilde{X}^{\mu}(\bar{z}) \rightarrow D^{\mu}_{\nu}X^{\nu}(\bar{z}) \ ,
\spa
\tilde{\psi}^{\mu}(\bar{z}) \rightarrow
D^{\mu}_{\nu}\psi^{\nu}(\bar{z}) \ ,
\spa
\tilde{\phi}(\bar{z}) \rightarrow \phi(\bar{z})\,, \mand
\tilde{S}_{\al}(\bar{z}) \rightarrow M_{\al}{}^{\be}{S}_{\be}(\bar{z})
 \ ,
\non\end{displaymath}

with 
\begin{displaymath}
D = \left( \begin{array}{cc}
-1_{9-p} & 0 \\
0 & 1_{p+1}
\end{array}
\right) \ ,\,\, \mand
M_p = \left\{\begin{array}{cc}\frac{\pm i}{(p+1)!}\ga^{a_{1}}\ga^{a_{2}}\ldots \ga^{a_{p+1}}
\eps_{a_{1}\ldots a_{p+1}}\\ \frac{\pm 1}{(p+1)!}\ga^{a_{1}}\ga^{a_{2}}\ldots \ga^{a_{p+1}}\ga_{11}
\eps_{a_{1}\ldots a_{p+1}} \end{array}\right.
\non\end{displaymath}

where $M_p$ has taken the values for even and odd $p$ accordingly. Keeping in mind above remarks,  now one can start to consider the known two point functions for   $X^{\mu},\psi^\mu, \phi$ fields  
\begin{eqnarray}
\lan X^{\mu}(z)X^{\nu}(w)\ran & = & -\frac{\alpha'}{2}\eta^{\mu\nu}\log(z-w) \ , \non \\
\lan \psi^{\mu}(z)\psi^{\nu}(w) \ran & = & -\frac{\alpha'}{2}\eta^{\mu\nu}(z-w)^{-1} \ ,\non \\
\lan\phi(z)\phi(w)\ran & = & -\log(z-w) \ .
\labell{prop2}\end{eqnarray}

It is discussed in \cite{Hatefi:2016yhb}  that to be able to generate all world volume contact terms of the amplitude of an RR and two real tachyons, the S-matrix produces all order $\alpha'$ corrections in both type II as follows

\beqa
i\mu_p (2\pi\alpha')^2  \int_{\Sigma_{p+1}} d^{p+1}\sigma \quad C_{(p-1)}\wedge \Tr\left(\sum_{m=-1}^{\infty}c_m(\alpha' (D ^b D_b))^{m+1}  DT \wedge DT\right) \labell{highaaw3}\eeqa

 Note that using S-matrix, one actually explores a new couplings for one RR, a gauge field and the scalar field (that describes transverse oscillations of the brane) of the form 
$(2\pi\alpha')^2\int d^{p+1} \sigma  \Tr( C^{i}_{p-2} \wedge F \wedge D \phi^i)$ where accordingly S-matrix generates its all order corrections as follows

  \beqa 
\frac{\mu_p (2\pi\alpha')^2}{(p-1)!}\int_{\sum_{(p+1)}}\Tr\bigg( C^{i}_{p-2} \wedge \sum_{n=-1}^{\infty}b_{n}(\alpha')^{n+1}  D_{a_{1}}\cdots D_{a_{n+1}} \wedge F D^{a_{1}}\cdots D^{a_{n+1}} \wedge D \phi^i\bigg)\labell{new28722}
 \eeqa
 
  \vskip.1in
  
Let us just denote the mixed Chern-Simons,  Taylor expansion or pull-back effective actions that are found by direct S-matrix computations.
\footnote{
\beqa
\frac{2i\beta'\mu_p'}{p!}(2\pi\alpha')^2 \int_{\Sigma_{p+1}} \partial_i C_{p}\wedge DT\phi^i \nonumber\\ 
\frac{2i\beta'\mu_p'}{p!}(2\pi\alpha')^2 \int_{\Sigma_{p+1}}  C^{i}_{p-1}\wedge DT\wedge D\phi^i \labell{highaa22}\eeqa

In the trace that involves  $\gamma^{11}$, we have special property  so that
\beqa
  p>3 , H_n=*H_{10-n} , n\geq 5.
  \nonumber\eeqa}
Having taken all the symmetries of the scattering amplitudes, one would also need to take into account some Bianchi identities that are restricted either on world volume or entire space-time including transverse directions. \footnote{
\beqa
  p_a H_{a_0...a_{p-1}}\epsilon^{a_0...a_{p-1}a}=0 \nonumber\\
  p^i \epsilon^{a_0...a_{p}}H_{a_0...a_{p}}+p^a \epsilon^{a_0...a_{p-1}a}H^{i}_{a_0...a_{p-1}}=0
  \nonumber\eeqa} Note that both the lower order DBI effective actions, Wess-Zumino (WZ) and  Chern-Simons couplings of D-brane Anti D-brane system  for general $D_p\bD_p$  of type II are explored  in section 2 of \cite{Hatefi:2017ags}. Given above information, we are now ready to write down the lower order  action that describes interactions between fermions and  tachyons. Let us take all of them into account and keep them fixed  into our lower order effective action, where we used the so called static gauge.
  
  \vskip.1in
   Hence, the Lagrangian is given by

\beqa
-T_pV(T)\sqrt{-\det(\eta_{ab}+2\pi\alpha'F_{ab}-2\pi\alpha'\bar\Psi\ga_b\prt_a\Psi+\pi^2\alpha'^2\bar\Psi\ga^{\mu}\prt_a\Psi\bar\Psi\ga_{\mu}\prt_b\Psi + 2\pi\alpha'\prt_aT\prt_bT )}
\label{esi111}
\eeqa


 \vskip .1in

Note that the tachyon potential that can be essentially given rise to consistent results  with superstring  calculations is revealed to be
 \beqa
V(|T|)&=&1+\pi\alpha'm^2|T|^2+
\frac{1}{2}(\pi\alpha'm^2|T|^2)^2+\cdots
\non\eeqa 
and $m_{T}^2=-1/(2\alpha')$ 
 which is very consistent with $V(|T|)=e^{\pi\alpha'm^2|T|^2}$  \cite{Kutasov:2000aq}. Notice that to be able to produce all the S-matrix elements of string theory, one must define  various new string couplings that include not only new structures but also their coefficients can be fixed out by S-matrix calculations. Those terms do have the structures such as $F^{(1)}\cdot{F^{(2)}}$ as well as $D\phi^{(1)}\cdot{D\phi^{(2)}}$ \cite{Hatefi:2016yhb}.

 \vskip .2in
 
 We also argued that the expansion of potential does work up to fourth order and generates consistent outcomes with string amplitudes, while once we send the tachyon to infinity the so called effective term  $T^4V(TT)$ goes to zero which is reasonable given the fact that we expect from a tachyon condensation of an unstable brane.

 \vskip .2in
 
 The so called  WZ action for D-brane Anti D-brane system  includes RR potentials just for $C=\sum_n(-i)^{\frac{p-m+1}{2}}C_m$ is \cite{Douglas:1995bn} and is shown by
\beqa
S = \mu_p\int_{\Sigma_{(p+1)}} C \wedge  \left(e^{i2\pi\alpha'F^{(1)}}-e^{i2\pi\alpha'F^{(2)}}\right)\ ,
\labell{eqn.wz}
\eeqa
It is also mentioned in  \cite{Kraus:2000nj} that one is able to replace the tachyons into the action while there is also another approach that can be 
 explained by the language of geometry and made use of super-connection formalism~\cite{quil,berl,Roepstorff:1998vh}. The following geometrical interpretations for WZ action are taken accordingly 
 \beqa
S_{WZ}&=&\mu_p \int_{\Sigma_{(p+1)}} C \wedge \STr e^{i2\pi\alpha'\cal F}\labell{WZ}\eeqa 
The definition of the curvature and connection is \beqa {\cal F}&=&d{\cal A}-i{\cal A}\wedge\cal A\nonumber \eeqa
and 
\begin{displaymath}
i{\cal A} = \left(
\begin{array}{cc}
  iA^{(1)} & \beta T^* \\ \beta T &   iA^{(2)} 
\end{array}
\right) \ ,
\non\end{displaymath}
where one finds the curvature out\footnote{ $F^{(i)}=\frac{1}{2}F^{(i)}_{ab}dx^{a}\wedge dx^{b}$ and $DT=[\partial_a T-i(A^{(1)}_{a}-A^{(2)}_{a})T]dx^{a}$}
\begin{displaymath}
i{\cal F} = \left(
\begin{array}{cc}
iF^{(1)} -\beta^2 |T|^2 & \beta (DT)^* \\
\beta DT & iF^{(2)} -\beta^2|T|^2 
\end{array}
\right) \ ,
\non\end{displaymath}
Now one is ready to explore all  the WZ couplings from  \reef{WZ} where we are going to consider just some of them that  have the relevant structures for the comparison with the EFT computations as follows \beqa
C\wedge \STr i{\cal F}&\!\!\!\!=\!\!\!&C_{p-1}\wedge(F^{(1)}-F^{(2)})\labell{exp2}\\
C\wedge \STr i{\cal F}\wedge i{\cal F}&\!\!\!\!=\!\!\!\!&C_{p-3}\wedge \left\{F^{(1)}\wedge F^{(1)}-
F^{(2)}\wedge F^{(2)}\right\}\nonumber\\
&& +C_{p-1}\wedge\left\{-2\beta^2|T|^2(F^{(1)}-F^{(2)})+2i\beta^2 DT\wedge(DT)^*\right\}\nonumber
\eeqa

From now on we concentrate on   $(C\bar \Psi_1\Psi_2 T_3T_4)$ amplitude to be able to find the higher order S-matrix elements of  D-brane  Anti D-brane system, and also start discovering all order their effective couplings as well as their corrections where the needed singularity structures will be obtained as well.

  \section{  All order S-matrix elements of $(C\bar \Psi_1\Psi_2 T_3T_4)$ }

As can be seen from \reef{esi111}, all order corrections for the non Abelian $\alpha'$ corrections to two tachyon-two fermion fields of particular brane Anti brane are not found out yet. Therefore, we carry out the direct computations of one RR, two fermions (with the same chirality) and two tachyons at disk level to discover all the desired non Abelian $\alpha'$ corrections of type II string theory.

 \vskip .2in

Note that it is of importance to highlight the fact that by just doing  $(C\bar \Psi_1\Psi_2 T_3T_4)$ amplitude, one is certainly able to exactly fix all order coefficients of all two tachyon two fermion couplings in the world volume of D-brane Anti D-brane system, because dualities are no longer  promising in this context.
   \vskip .2in
   
We  would like to deal with the S-matrix elements of a closed string RR, two fermion fields and two real tachyons of D-brane Anti D-brane system to be able to construct higher order effective actions of this particular system of interest. Basically using all the CFT methods, we look for 
  $<V_{C^{-1}(z,\bar z)}  V_{\bar\psi^{-1/2}(x_1)}  V_{\psi^{-1/2}(x_2)} V_{T^{0}(x_3)} V_{T^{0}(x_4)}>$
 S-matrix. We also try to explore the correct momentum expansion of the S-matrix and ultimately try to regenerate all singularity structures of  type II superstring theory where  all the symmetries of the string amplitude and  a special soft limit are taken into account. Note that given the world volume structures in this paper, we no longer have any bulk poles to carry momentum of RR in transverse direction.  
 
 \vskip.2in
 
Having taken into account all the described vertex operators, one would search for the correlation function of four spin operators and two world sheet fermions for which the following correlation needs to be derived \footnote{ To make the correlators singlet we need to consider two fermion fields with the same chirality, for further information see \cite{Hatefi:2013mwa}.}

 \beqa
I_{\al\be\ga\delta}^{ab}&=&<:S_{\al}(z): S_{\be}({\bar z}):S_{\ga}(x_1): S_{\delta}({x_2}):\psi^{a}(x_{3}):\psi^{b}(x_{4}):>\label{88az}\eeqa
  
 In section 2 of \cite{Hatefi:2013hca} and also in section 2.1 of \cite{Hatefi:2014saa} it is explained how to derive the correlation function between four spin operators an one current where essentially the same method can be used to explore the above correlation function, however, let us briefly talk about its derivation. To be able to carry this out, one needs to first take the OPE between one spin operator and all the fermion fields where all permutations need to be regarded and then one needs to sum them up. Finally,  one has to substitute the outcome to the primary correlation which would be the correlation function between four spin operators that is given in section 2 of \cite{Hatefi:2013hca} and after all we start constructing all possible ways of different symmetric and antisymmetric gamma matrices. Eventually the closed form of $I_{\al\be\ga\delta}^{ab}$ can be given by       
  \beqa
I^{ba}_{\alpha\beta\gamma\delta}&=&\bigg[\frac{\eta^{ab}}{x_{34}} (\gamma^{\lambda}C)_{{\alpha\beta}}(\gamma_{\lambda }C)_{{AB}}x_{52}  x_{61}(x_{35}  x_{32}  x_{46}  x_{41}+x_{36}  x_{31}  x_{45}  x_{42})\nonumber\\&&-\frac{\eta^{ab}}{x_{34}} (\gamma^{\lambda }C)_{{\alpha B}}(\gamma_{\lambda }C)_{{A\beta}}x_{56}  x_{12}(x_{35}  x_{36}  x_{41}  x_{42}+x_{31}  x_{32}  x_{45}  x_{46})\nonumber\\&&
+\frac{1}{2} (\gamma^{a }C)_{{A\beta}}(\gamma_{b }C)_{{\alpha B}}x_{56}  x_{12}(x_{35}  x_{41}  x_{62} -x_{32}  x_{46}  x_{51})\nonumber\\&&+\frac{1}{2} (\gamma^{a }C)_{{\alpha B}}(\gamma_{b }C)_{{A\beta}}x_{56}  x_{12}(x_{36}  x_{42}  x_{51} -x_{31}  x_{45}  x_{62})
\nonumber\\&&-\frac{1}{2} (\gamma^{a }C)_{{\alpha \beta}}(\gamma_{b }C)_{{AB}}x_{52}  x_{61}(x_{32}  x_{46}  x_{51} +x_{31}  x_{45}  x_{62})\nonumber\\&&+\frac{1}{2} (\gamma^{a }C)_{{AB}}(\gamma_{b }C)_{{\alpha \beta}}x_{52}  x_{61}(x_{35}  x_{41}  x_{62} +x_{36}  x_{42}  x_{51})\nonumber\\&&+\frac{1}{4} (\Gamma^{ab\lambda }C)_{{\alpha \beta}}(\gamma_{\lambda }C)_{{AB}}x_{56}x_{52}  x_{61}(x_{31}  x_{42}+x_{32}  x_{41})\nonumber\\&&+\frac{1}{4} (\Gamma^{ab\lambda }C)_{{AB}}(\gamma_{\lambda }C)_{{\alpha \beta}}x_{52}x_{61}  x_{12}(x_{35}  x_{46}+x_{36}  x_{45})\nonumber\\&&-\frac{1}{4} (\Gamma^{ab\lambda }C)_{{\alpha B}}(\gamma_{\lambda }C)_{{A \beta}}x_{56}x_{52}  x_{12}(x_{31}  x_{46}+x_{36}  x_{41})\nonumber\\&&+\frac{1}{4} (\Gamma^{ab\lambda }C)_{{A\beta}}(\gamma_{\lambda }C)_{{\alpha B}}x_{56}x_{61}  x_{12}(x_{35}  x_{42}+x_{32}  x_{45})\nonumber\\&&- (\gamma^{a }C)_{{\alpha A}}(\gamma^{b }C)_{{\beta B}}x_{34}x_{56}  x_{52}x_{61}  x_{12}\bigg]\frac{(x_{56}  x_{51}  x_{52}  x_{61}  x_{62}  x_{12})^{-3/4}}{2 (x_{35}  x_{36}  x_{31}  x_{32}x_{45}  x_{46}  x_{41}  x_{42})^{1/2}}
\nonumber\eeqa
where  $x_5=z=x+iy, x_6=\bar z$. The method of finding  a six point bosonic amplitude has been recently announced in \cite{Hatefi:2017ags}, essentially one can use the same methodology for fermionic amplitudes  as well. One finds out the closed and final form of the amplitude as below

 \beqa
{\cal A}^{C\bar\psi \psi TT}&\sim& 2\int dx_{1}dx_{2}
dx_{3}dx_{4}dx_{5}dx_{6}(P_{-}\fsH_{(n)}M_p)^{\al\be}  ( x_{12}x_{15}x_{16}x_{56})^{-1/4}(x_{25}x_{26})^{-1/4}\nonumber\\&&\times
\bar u^A u^B (-\alpha'^2k_{3a}k_{4b}) I_{\al\be\ga\delta}^{ab} I,\eeqa
where
 
 \beqa
I&=&|x_{12}|^{-2t}|x_{13}|^{-2s-\frac{1}{2}}|x_{14}|^{-2v-\frac{1}{2}}|x_{23}|^{-2u-\frac{1}{2}}
|x_{24}|^{-2r-\frac{1}{2}}|x_{34}|^{-2w-1}|x_{15}x_{16}|^{t+s+v+\frac{1}{2}} \nonumber\\&&\times
|x_{25}x_{26}|^{t+u+r+\frac{1}{2}}|x_{35}x_{36}|^{s+u+w+\frac{1}{2}}|x_{45}x_{46}|^{v+r+w+\frac{1}{2}}|x_{56}|^{-2(s+t+u+v+r+w)-2}
\nonumber\eeqa
 
Now the SL(2,R) invariance of the amplitude can be obviously notified.  We now try to do gauge fixing by fixing the position of open strings at $x_1=0, 0\leq x_2\leq 1 , x_3=1, x_4=\infty$ and especially we try to introduce six Mandelstam variables as $s=-(\frac{1}{4}+2k_1.k_3), t=-(2k_1.k_2), v=-(\frac{1}{4}+2k_1.k_4), u=-(\frac{1}{4}+2k_2.k_3), r=-(\frac{1}{4}+2k_2.k_4), w=-(\frac{1}{2}+2k_3.k_4)$ where setting this sort of gauge fixing we would obtain the amplitude as follows

 \beqa 
  &&
\int_{0}^{1} dx_2 x_2^{-2t-1} (1-x_2)^{-2u-1}\int dz \int d\bar z |1-z|^{2s+2u+2w} |z|^{2t+2s+2v-1}
 (z - \bar{z})^{-2(t+s+u+v+r+w)-3}  \nonumber\\&&\times|x_2-z|^{2t+2u+2r-1}
 \bar u^A u^B (-\alpha'^2k_{3a}k_{4b}) (P_{-}\fsH_{(n)}M_p)^{\alpha\beta} \nonumber\\&&\times
\bigg[-\eta^{ab} (\gamma^{\lambda}C)_{{\alpha\beta}}(\gamma_{\lambda }C)_{{AB}} (z-x_2) \bar z( (1-z)(1-x_2)+(1-\bar z))
\nonumber\\&&-\eta^{ab} (\gamma^{\lambda }C)_{{\alpha B}}(\gamma_{\lambda }C)_{{A\beta}} (z-\bar z) x_2(|1-z|^2+(1-x_2))
\nonumber\\&&
-\frac{1}{2} (\gamma^{a }C)_{{A\beta}}(\gamma_{b }C)_{{\alpha B}} (z-\bar z) x_2 ((1-z)(\bar z-x_2)-(1-x_2)z)
\nonumber\\&&-\frac{1}{2} (\gamma^{a }C)_{{\alpha B}}(\gamma_{b }C)_{{A\beta}} (z-\bar z) x_2 ((1-\bar z)z-(\bar z-x_2))
\nonumber\\&&-\frac{1}{2} (\gamma^{a }C)_{{\alpha \beta}}(\gamma_{b }C)_{{AB}} (z-x_2)\bar z ((1-x_2) z+(\bar z-x_2))
\nonumber\\&&+\frac{1}{2} (\gamma^{a }C)_{{AB}}(\gamma_{b }C)_{{\alpha \beta}} (z-x_2)\bar z ((1-z) (\bar z-x_2)+(1-\bar z) z)
\nonumber\\&&+\frac{1}{4} (\Gamma^{ab\lambda }C)_{{\alpha \beta}}(\gamma_{\lambda }C)_{{AB}} (z-\bar z) (z-x_2) \bar z (1+(1-x_2))
\nonumber\\&&-\frac{1}{4} (\Gamma^{ab\lambda }C)_{{AB}}(\gamma_{\lambda }C)_{{\alpha \beta}}(z-x_2) \bar z x_2 ((1-z)+(1-\bar z))
\nonumber\\&&
+\frac{1}{4} (\Gamma^{ab\lambda }C)_{{\alpha B}}(\gamma_{\lambda }C)_{{A \beta}} (z-\bar z)(z-x_2) x_2(1+(1-\bar z))
\nonumber\\&&-\frac{1}{4} (\Gamma^{ab\lambda }C)_{{A\beta}}(\gamma_{\lambda }C)_{{\alpha B}} (z-\bar z) (\bar z)  x_2   ((1-z)+(1-x_2))\nonumber\\&&- (\gamma^{a }C)_{{\alpha A}}(\gamma^{b }C)_{{\beta B}} (z-\bar z) (z-x_2) (\bar z) x_2\bigg] \label{ee66}\eeqa
One finds out that the whole amplitude has non-vanishing contributions for  $C_{p-3},C_{p-1},C_{p+1}$ cases. As has been argued in \cite{Hatefi:2017ags} one can immediately explore the algebraic 
solutions for the entire integrals in the soft limit  where they get decoupled. \footnote{We take $4k_2.p \rightarrow 2$ limit  which is shifted with a constant due to the fact that here we are dealing with fermionic couplings in the presence of RR and we no longer work with bosonic strings.} Basically the whole integrals are appeared in  \cite{Hatefi:2012wj} and \cite{Fotopoulos:2001pt} so one can write down the final answer of the amplitude in terms of the Gamma functions. For instance one derives the result for just the second term of \reef{ee66} as follows 

\beqa
 {\cal A}_{2}\sim\  2^{-2(t+s+u+v+r+w)-2} \bar u^A u^B (-2u-\frac{1}{2}) \frac{\Gamma(-2t+1) \Gamma(-2u)}{\Gamma(-2t-2u+1)}(P_{-}\fsH_{(n)}M_p)^{\alpha\beta} 
(\gamma^{\lambda}C)_{{\alpha B}}(\gamma_{\lambda }C)_{{A\beta}}  L_1\nonumber\eeqa

 \beqa
    L_1&=& \frac{\Gamma \left(r-s-\frac{1}{2}\right) \Gamma (-r-t-v) \Gamma \left(-r-u-w-\frac{1}{2}\right)  \Gamma \left(-r-s-t-u-v-w-\frac{1}{2}\right)}{\Gamma \left(-s-t-v+\frac{1}{2}\right) \Gamma (-s-u-w-1) \Gamma \left(-2 r-t-u-v-w-\frac{1}{2}\right)} \nonumber\\&&\times 
  \left(\frac{u (-2 r+2 s+1)}{(2 t+2 u-1) (s+u+w+1)}+\frac{2 (r+t+v)}{4 r+2 t+2 u+2 v+2 w+1}\right)   \eeqa

We have already followed out the following papers  \cite{Hatefi:2012wj} and \cite{Fotopoulos:2001pt} to be able to derive all the other parts of the amplitude, however, due to the fact that the final solutions for the integrals are tedious and cumbersome we decided not to write down the entire terms but essentially the procedure would be the same as we prescribed for the second term of the amplitude. Now the non trivial question is related to the expansion of the amplitude  which we talk about it in the next section.

\section{Momentum expansion of $D_{p} \bar D_{p}$ }

In order to explore all the singularity structures, one needs to expand out the amplitude. Having set out the remarks given in \cite{Hatefi:2017ags}, we would highlight some further points that are relevant to the expansion of the fermionic amplitudes. Taking momentum conservation along the world volume of brane,  one gets to derive
 \beqa
  s+t+u+v+r+w=-p^a p_a-1\label{expabc}\eeqa
 It is discussed that for D-brane Anti D-brane system, the proper expansion can be found by sending the quantity $p^a p_a$ to zero while for non-BPS branes and making use of a two point $CT$ amplitude one finds that, this quantity must tend to mass of the tachyon.  The proper expansion is largely discussed in \cite{Hatefi:2012wj}. We just illustrate the fact that there is  a non-zero coupling between two fermions and a gauge field, hence $t\rightarrow 0$ and there are no two fermions and one tachyon couplings, however since $k_i.k_j$ must tend to zero, given the definition of Mandelstam variables we come to know  that the other four Mandelstam variables  should be sent to mass of tachyon, namely $s,v,u,r\rightarrow \frac{-1}{2\alpha'} $. 
 
  \vskip.2in
 
 Now given \reef{expabc},  one reveals $w$ must also be sent to zero where it is also consistent with $CTT$ amplitude. Hence we clearly found out the proper momentum expansion of the amplitude  by making use of the symmetries of the amplitude as well as  insisting on  all EFT parts. Hence, the momentum expansion is summarised in below 
 
  \beqa
\bigg(s,v,u,r\rightarrow \frac{-1}{2\alpha'} ,    \quad t,w\rightarrow 0\bigg), \label{expas111}
\eeqa

As an example one can show that the amplitude involves t-channel gauge field singularities  for $p=n+2$ case as well as an infinite tachyon singularities in $(t+s+u+\frac{1}{2})$ and $(t+v+r+\frac{1}{2})$ cases for  $p=n$ case accordingly. Due to symmetries we just produce all infinite $(t+s+u+\frac{1}{2})$ channel tachyonic poles by constructing all infinite higher derivative $\alpha'$ corrections to two tachyons and two fermion fields in the world volume of D-brane Anti D-brane system. We also produce all t-channel singularities in an EFT sub amplitude later on. 

\vskip.1in
  
By dealing with EFT methods and also considering the symmetries of the amplitude \cite{Schwarz:2013wra}, more crucially relying on  selection rules \cite{Hatefi:2013yxa}, we come to understand that  there is no interaction between two fermions and a tachyon nor two tachyons and scalar field , therefore we expect not to have $s',v',u',r'$ channel poles whatsoever.

  \vskip.1in
  
All the kinetic terms of fermionic strings, gauge fields and tachyons have already been fixed in the DBI action as follows

\beqa
-T_p \frac{2\pi\alpha'}{2}
  \Tr\left( \bar \Psi\gamma ^a D_a\Psi + D_aTD^aT-\frac{(2\pi\alpha')}{2}F_{ab}F^{ba}\right)\labell{kinetic terms}
\eeqa

where
\beqa
D_a\Psi &=&\partial_a\Psi-i[A^a,\Psi], \quad D_aT=\partial_a T-i[A^a,T], \nonumber\eeqa

Let us now explore  all infinite $\alpha'$ corrections to  $\bar\Psi \Psi TT$ couplings in the world volume of D-brane Anti D-brane system.

\section {All $\alpha'$ Corrections to $\bar\Psi \Psi T T$  Couplings of $D_{p} \bar {D_p}$}
\vskip.2in

In order to derive all infinite tachyon singularities of the amplitude for $p=n$ case, one needs to know all infinite $\alpha'$ higher derivative corrections to two fermion-two tachyon couplings in the world volume of D-brane Anti D-brane system, which we are going to obtain them accordingly.  First of all let us comment on lower point functions of string amplitudes. The Wess-Zumino action  of an RR and a tachyon is shown by 
\beqa
2i\beta\mu'_p(2\pi\alpha')\int_{\Sigma_{p+1}} C_p\wedge DT
\labell{d22}\eeqa

where it is shown that \reef{d22} receives no correction and also the vertex  of an RR and two tachyons is fixed, note that  the kinetic term of tachyon in DBI action has also been fixed, hence tachyon propagator does not receive any correction either. Thus to be able to generate all tachyon singularities one needs to apply the correct higher derivative corrections to  $\bar\Psi \Psi T T$  effective couplings of $D_{p} \bar {D_p}$  as follows.

\vskip.1in

In order to find the coefficients of two tachyon-two fermion couplings, one needs to first deal with pure open string calculations of $\bar\Psi \Psi T T$ where detailed explanation was given  in \cite{Hatefi:2013mwa}, however, for the sake of clarity, here we summarise some of the basic results.

    \vskip .2in

The ultimate form of the amplitude of $\bar\Psi \Psi T T$ must be symmetric with respect to exchanging tachyons and also should be  anti symmetric under interchanging two fermion fields. To restore all the symmetries we find out all the possible six orderings of the open strings. For instance just for $\Tr(\lam_1\lam_2\lam_3\lam_4)$ ordering we obtain

\beqa
{\cal A}^{\bar \Psi,\Psi,T,T}  &= & (-8 T_p)\Tr(\sigma_3 I\sigma_2\sigma_1) \bar u_1^A(\ga^{a})_{AB}u_2^B (\alpha'ik_{4a})\int_0^{1} dx_2 x_2^{-2t-1}(1-x_2)^{-2u-1} \Tr(\lam_1\lam_2\lam_3\lam_4)
\nonumber\\ & =&(-8 T_p) \bar u_1^A(\ga^{a})_{AB}u_2^B (\alpha'ik_{4a}) \Tr(\lam_1\lam_2\lam_3\lam_4)\Tr(\sigma_3 I\sigma_2\sigma_1)\frac{\Gamma(-2t)\Gamma(-2u)}{\Gamma(-2t-2u)}\eeqa

The ultimate form of the amplitude is given by
\beqa
{\cal A}^{\bar \Psi,\Psi,T,T} & = &(-8 T_p ) \bar u_1^A(\ga^{a}_{AB})u_2^B (\alpha'ik_{4a})\Tr(\sigma_3 I\sigma_2\sigma_1)\bigg(l_1\frac{\Gamma(-2t)\Gamma(-2u)}{\Gamma(-2t-2u)}-l_2 \frac{\Gamma(-2t)\Gamma(-2s)}{\Gamma(-2t-2s)}\nonumber\\&&
+il_3\frac{\Gamma(-2u)\Gamma(-2s)}{\Gamma(-2u-2s)}\bigg)\labell{amp1}
\eeqa
where $l_1,l_2,l_3$ are defined in \cite{Hatefi:2013mwa}, given the on-shell condition  $s+t+u=-\frac{1}{2}$ and the fact that we know there is a gauge field pole and a non zero coupling between two fermions and a gauge field also due to symmetries under exchanging two tachyons, we get to know that the expansion must be around 
\beqa
t\rightarrow 0,\qquad s,u\rightarrow -\frac{1}{4}\label{143e}
\eeqa
Defining  $u'=u+1/4=-\alpha'k_2\inn k_3$ and $s'=s+1/4=-\alpha'k_1\inn k_3$, the on-shell condition becomes $s'+t+u'=0$ and the amplitude gets changed to
\beqa
{\cal A}^{\bar \Psi_1,\Psi_2,T_3,T_4} & \!=\! &(-8 T_p )\bar u_1^A(\ga^{a}_{AB})u_2^B (\alpha'ik_{4a})\Tr(\sigma_3 I\sigma_2\sigma_1)
\bigg(l_1\frac{\Gamma(2u'+2s')\Gamma(\frac{1}{2}-2u')}{\Gamma(\frac{1}{2}+2s')}\nonumber\\&&-l_2 \frac{\Gamma(2u'+2s')\Gamma(\frac{1}{2}-2s')}{\Gamma(\frac{1}{2}+2u')}+il_3\frac{\Gamma(\frac{1}{2}-2s')\Gamma(\frac{1}{2}-2u')}{\Gamma(1-2s'-2u')}
\bigg)\labell{amp4}
\eeqa

\vskip.2in

Expanding around \reef{143e}, one reads off the amplitude as 
\beqa
{\cal A}^{\bar \Psi_1,\Psi_2,T_3,T_4} & = &-8 T_p
  \bar u_1^A(\ga^{a}_{AB})u_2^B (\alpha'ik_{4a})\Tr(\sigma_3 I\sigma_2\sigma_1)    \labell{amp52}\\&&
  \bigg(\frac{l_1-l_2}{-2t}+\sum_{n,m=0}^{\infty}\bigg[a_{n,m}(l_1 u'^ns'^m-l_2 s'^n u'^m)+il_3 b_{n,m}(s'^nu'^m+ s'^mu'^n)\bigg]\bigg)\nonumber\eeqa

\vskip.2in

If we use \reef{143e} then one reads off all the coefficients of $a_{n,m},b_{n,m}$ in below
\beqa
&&a_{0,0}=2\ln(2),b_{0,0}=\frac{\pi}{2},a_{1,1}=-12\z(3)+\frac{32}{3}\ln(2)^3+\frac{4\pi^2}{3}\ln(2),\,b_{1,1}=\frac{\pi}{3}(-\pi^2+24\ln(2)^2),\nonumber\\
&&a_{1,0}=\frac{2\pi^2}{3}+4\ln(2)^2,\,b_{1,0}=2\pi\ln(2),\,a_{0,1}=-\frac{\pi^2}{3}+4\ln(2)^2,\nonumber\\
&&a_{2,0}=8\z(3)+\frac{16}{3}\ln(2)^3+\frac{8\pi^2}{3}\ln(2),b_{2,0}=\frac{\pi}{3}(\pi^2+12\ln(2)^2)\nonumber\\
&&\,a_{0,2}=8\z(3)+\frac{16}{3}\ln(2)^3-\frac{4\pi^2}{3}\ln(2),\,b_{1,2}=\frac{4\pi}{3}(12\ln(2)^3-3\z(3)).\label{coefs}
\eeqa

where all $b_{n,m}$ coefficients are indeed symmetric. It is shown that the t- channel gauge field pole is generated by a sub-amplitude in an EFT part. \footnote{
$V^{a,\alpha}(\bar \Psi_1,\Psi_2, A)G^{\alpha\beta}_{ab}(A)V^{b,\beta}(A,T_3,T_4)$.}

\vskip.2in

 Contact interactions of the $\bar \Psi_1\Psi_2 T_3 T_4$ at zeroth order  can be found by
\beqa
{\cal A}^{\bar \Psi_1,\Psi_2,T_3,T_4} & = &-8 T_p (2i)
  \bar u_1^A(\ga^{a}_{AB})u_2^B (\alpha'ik_{4a}) \left[a_{0,0}(l_1 -l_2 )+2il_3 b_{0,0}\right]\nonumber
\eeqa
which should be at the same order of the kinetic term of fermions.\footnote{  $(k_1+k_2+k_3+k_4)^a=0 $  and on shell conditions $\gamma^a k_{1a} \bar u^A=\gamma^a k_{2a}  u^B=0 $ are also taken.} 

Now we can write down all possible couplings of two fermions two tachyons at first order in $\alpha'$ as follows

\beqa
\alpha'T_p\Tr\left(\frac{}{}a_{0,0}\left(\bar \Psi\ga^a\Psi TD_a T-\bar\Psi\ga^a\Psi D_aT T\right)+ib_{0,0}\left(\bar\Psi\ga^a T\Psi D_a T-\bar\Psi\ga^aD_aT\Psi T\right)\right)\labell{nonabcoup}
\eeqa

\vskip.2in 
If we use $(a_{0,1}-a_{1,0})$ term and for Abelian gauge group, the first term
in \reef{amp52} is reconstructed by considering the coupling $(2\pi^2\alpha'^2 T_p\bar \Psi\gamma^b \partial^a\Psi\partial_bT\partial_a T)$.

\vskip.2in

Now we would like to reconstruct the entire all order higher derivative corrections to two fermions and two tachyon couplings of D-brane Anti D-brane system. Indeed we are looking for non- Abelian interactions of $C \bar \Psi_1\Psi_2T_3T_4$. Note that some further remarks are given in section  six of \cite{Hatefi:2016yhb} for all order corrections of two scalar two tachyon couplings of $D_p \bar D_p$. \footnote{ We keep track of the over all coefficient of $
  \bar u_1^A(\ga^{a}_{AB})u_2^B ( i\alpha' k_{3a}) $ for all the terms in the ultimate form of the amplitude of $C \bar \Psi_1\Psi_2T_3T_4$ for $p=n$ case.} The  prescription for obtaining  all order higher derivative corrections are provided in \cite{Hatefi:2012rx,Hatefi:2012wj} accordingly, hence we first write down all order higher derivative corrections to non-Abelian couplings for two fermion-two tachyon effective couplings just in the world volume of D-brane Anti-D-brane system and then using them we try to reconstruct 
  all the tachyon singularities for $p=n$ case as well.  Hence, the Lagrangian is given by
 
  \beqa
{\cal L}^{n,m}&=&\frac{1}{2}\alpha'^{n+m+1}T_p\left(a_{n,m}\Tr\left[\frac{}{}\cD_{nm}\left(\bar\Psi^{(1)}\ga^a\Psi^{(1)} T D_a T^*\right)+\cD_{nm}\left(T D_a T^*\bar\Psi^{(1)}\ga^a\Psi^{(1)} \right)\right.\right.\nonumber\\
&&\left.\left.-\frac{}{}\cD_{nm}\left(\bar\Psi^{(1)}\ga^a\Psi^{(1)} D_aT  T^*\right)-\cD_{nm}\left(D_aT  T^*\bar\Psi^{(1)}\ga^a\Psi^{(1)} \right)+h.c\right]\right.\nonumber\\
&&\left.-ib_{n,m}\Tr \left[\cD'_{nm}\left(\bar\Psi^{(2)}\ga^a T\Psi^{(1)} D_a T^*\right)+\cD'_{nm}\left(\Psi^{(1)} D_aT\bar\Psi^{(2)} \ga^a  T^*\right)\right.\right.\nonumber\\
&&\left.\left.-\cD'_{nm}\left(\bar\Psi^{(2)}\ga^a D_aT\Psi^{(1)}  T^*\right)-\cD'_{nm}\left(\Psi^{(1)} T\bar\Psi^{(2)} \ga^a  D_aT^*\right)+h.c\right]\right)\labell{Lnm}
\eeqa

\vskip.2in

\vskip.2in

where we have defined \footnote{ The definition of special higher derivative operators $\cD_{nm}$ and $\cD'_{nm}$ is 
\beqa
\cD_{nm}(EFGH)&\equiv&D_{b_1}\cdots D_{b_m}D_{a_1}\cdots D_{a_n}E  F D^{a_1}\cdots D^{a_n}GD^{b_1}\cdots D^{b_m}H\nonumber\\
\cD'_{nm}(EFGH)&\equiv&D_{b_1}\cdots D_{b_m}D_{a_1}\cdots D_{a_n}E   D^{a_1}\cdots D^{a_n}F G D^{b_1}\cdots D^{b_m}H\nonumber
\eeqa} and $T=\frac{1}{2}(T_1+iT_2)$ and $T^*=\frac{1}{2}(T_1-iT_2)$ for which  $\bar\Psi^{(1)}$  is the massless fermion that lives on D-brane and  $\Psi^{(2)}$ is the massless fermion that lives on Anti D-brane and vice versa. It is worth mentioning the fact that the above couplings are indeed non-Abelian extensions of $(\alpha')^{1+n+m}$  order for   $a_{0,0}$ and $b_{0,0}$ coefficients.


  \subsection{All Infinite Tachyon Singularities  of $C\bar\Psi\Psi TT$ }

\vskip.2in

Having derived all new couplings with their exact coefficients in \reef{Lnm} in the world volume of D-brane Anti D-brane system, which include some new structures such as 

\beqa
\bigg(\Psi^{(1)} D_aT\bar\Psi^{(2)} \ga^a  T^*\bigg), \quad \quad 
\bigg(\Psi^{(1)} T\bar\Psi^{(2)} \ga^a  D_aT^*\bigg)\label{88}\eeqa

   we now turn to construct an infinite tachyon singularities of 
   the amplitude  $C\bar\Psi\Psi TT$ for $p=n$ case in type II string theory.
    Indeed we need to make use of  \reef{Lnm}, to be able to precisely generate all singularities of the amplitude. \footnote{Note that using direct computations at each order we have also shown that the couplings that are found for non-BPS  amplitudes  \cite{Hatefi:2012cp} are not applicable to the world volume of D-brane Anti D-brane system.}
\vskip.2in

The amplitude has tachyon singularities in $(t+s+u+\frac{1}{2})$ and $(t+v+r+\frac{1}{2})$ where due to symmetries we just try to produce just all tachyon singularities in $(t+s+u+\frac{1}{2})$ channels. If we extract the traces in the string amplitude and working out some further simplifications, one finds out all the tachyon singularities of the string amplitude  as follows

   \beqa
{\cal A}^{C\bar\Psi_1\Psi_2 T_3T_4}&=&\frac{8 \pi^2\mu'_p\beta}{p!}  \bar u_1^A(\ga^{a}_{AB})u_2^B
 \eps^{a_{0}\cdots a_{p-1}b}H_{a_{0}\cdots a_{p-1}} \Tr(\lam_1\lam_2\lam_3\lam_4)\nonumber\\&&\times
(2i\alpha' k_{3a}k_{4b})\sum_{n,m=0}^{\infty}\frac{e_{n,m}[s'^nu'^m -u'^ns'^m]}{(t+s'+u')}\label{nn89}\eeqa

  where $s'=s+\frac{1}{4}, u'=u+\frac{1}{4} $ and a normalisation constant $\frac{i\beta\mu'_p}{(2\pi)^{1/2}}$ is used. Note that $\mu'_p$ is  RR brane's charge and $\beta$  is called the Wess-Zumino normalisation constant ( see \cite{Hatefi:2016yhb} for further details).

We need to expand the amplitude so that all the singularities can be evidently reproduced and to do so we use the momentum expansion that we talked about and also further ingredients about the expansion of fermionic amplitudes can be taken from section 2 and 3 of \cite{Hatefi:2013mwa}.   Simultaneously, we keep track of  on shell conditions for fermions as well as momentum conservation along the  brane.
   
   \vskip.1in
 
  Consider the following sub amplitude in an effective field theory to be able to generate all tachyon singularities
 
  \beqa
{\cal A}&=&V^{\alpha}(C_{p-1},T_4,T)G^{\alpha\beta}(T)V^{\beta}(T,T_3,\bar\Psi_1,\Psi_2)\labell{amp5441}\eeqa
 with the following vertices

\beqa
G^{\alpha\beta}(T) &=&\frac{i\delta^{\alpha\beta}}{(2\pi\alpha') T_p
(k^2+m^2)}\nonumber\\
V^{\alpha}(C_{p-1},T_3,T)&=&i\mu'_p\beta(2\pi\alpha')^2\frac{1}{p!}\epsilon^{a_0\cdots a_{p}}H_{a_0\cdots a_{p-1}}k_{4_{a_p}}\Tr(\lambda^{4}\lambda^{\alpha})
\labell{Fey1}
\eeqa

\vskip.1in

 where  the Chern-Simons coupling $\mu'_p \beta(2\pi\alpha')^2 \int_{\Sigma_{p+1}} d^{p+1} \sigma \quad C_{p-1}\wedge DT\wedge DT$ has been employed. The propagator can also be read off as $\frac{i\delta^{\alpha\beta}}{(2\pi\alpha') T_p
(t+s'+u')}$. 

\vskip.1in

Now we try to explore all order generalisation of 
 the  vertex of $ V^{\beta}(T,T_3,\bar\Psi_1,\Psi_2)$ by employing the all order corrections of two fermion two tachyon couplings of the discovered higher derivative non-abelian couplings that are constructed in \reef{Lnm}.

\vskip.1in

  In order to derive all oder generalisation of  $V^{\beta}(T,T_3,\bar\Psi_1,\Psi_2)$ where two fermions are on-shell while one tachyon is off-shell and the other tachyon that we labelled as $T_3$ is an external tachyon, we need to find the related vertex from an EFT. Let us consider the following ordering  $\Tr(\lambda_1\lambda_2\lambda_3\lambda_{\beta})$ where $\beta$ is related to the Abelian tachyon and also for the sake of simplicity we just concentrate on the terms of the Lagrangian with $ a_{n,m}$ coefficients. For instance if we consider

\beqa\cD_{nm}\left(\bar\Psi^{(1)}\ga^a\Psi^{(1)} T D_a T^*\right)+\cD_{nm}\left(T D_a T^*\bar\Psi^{(1)}\ga^a\Psi^{(1)} \right)  \nonumber\eeqa

we then can derive the following vertices for the above effective couplings

\beqa
a_{n,m}(k.k_1)^m (k_1.k_3)^n  \bar u\ga^a u (-ik_{4a})\nonumber\\
+a_{n,m}(k_3.k_2)^m (k_3.k_1)^n \bar u\ga^a u (-ik_{4a})
\eeqa
and one can do the same for the other terms in the Lagrangian, basically for the following  terms 
 \beqa
  -\cD_{nm}\left(\bar\Psi^{(1)}\ga^a\Psi^{(1)} D_aT  T^*\right)-\cD_{nm}\left(D_aT  T^*\bar\Psi^{(1)}\ga^a\Psi^{(1)} \right)\nonumber
\eeqa
  
we find out the following vertices  accordingly
\beqa
-a_{n,m}(k_1.k)^m (k_1.k_3)^n \bar u\ga^a u (-ik_{3a})\nonumber\\
-a_{n,m}(k_3.k_2)^m (k_3.k_1)^n \bar u\ga^a u (-ik_{3a})
\eeqa

 where $k$ becomes the momentum of  off-shell tachyon.  We need to consider the other ordering, namely since the tachyon propagator is off-shell we need to consider $\Tr(\lambda_1\lambda_2\lambda_{\beta}\lambda_3)$ ordering as well and also take into account the hermitian conjugate of the other terms in \reef{Lnm}, then one is able to explore the ultimate form of vertex in an EFT for $a_{n,m}$. The same is applied for $b_{n,m}$ as well. Now if we use momentum conservation  along the brane $(k_1+k_2+k_3+k+4)_a=0$ and also take into account the on-shell condition for fermion fields as  $\ga^a k_{1a} \bar u^A=\ga^a k_{2a} u^B=0$ then we are able to obtain the ultimate form of $V^{\beta}(T,T_3,\bar\Psi_1,\Psi_2)$ as follows
    \beqa
   V_{\beta}^{b}(\bar\Psi_1,\Psi_2,T_3,T)&=& iT_{p} (\alpha')^{n+m+1}(a_{n,m}-ib_{n,m})
\bar u_1^{A}(\ga^a)_{AB} u_2^{B}  (2ik_{3a})\nonumber\\&&
\bigg[-(k_1.k)^m (k_3.k_1)^n+(k_1.k)^n (k_3.k_1)^m+(k.k_2)^m (k.k_1)^n\nonumber\\&&
-(k.k_2)^n (k.k_1)^m
+(k_3.k_1)^m(k_2.k_3)^n-(k_3.k_1)^n(k_3.k_2)^m\nonumber\\&&
+ (k.k_2)^m (k_2.k_3)^n-(k.k_2)^n (k_2.k_3)^m
\bigg]
  \Tr(\lam_1\lam_2\lam_3\lambda_{\beta})
 \labell{verpt}\eeqa

If we replace the above vertex operator inside the sub amplitude in the EFT part as indicated in \reef{amp5441}, then one discovers all infinite tachyon singularities in below :

  \beqa
&&32 i k_{3a}\beta\mu'_p  \eps^{a_{0} \cdots a_{p}}  H_{a_0\cdots a_{p-1}} k_{4a_{p}}  \frac{1}{p!(s'+t+u')}
\nonumber\\&&\times
\sum_{n,m=0}^{\infty} (a_{n,m}-ib_{n,m})[s'^{m}u'^{n}-s'^{n}u'^{m}]
     \bar u_1^{A} (\ga^a)_{AB} u_2^B
   \Tr(\lam_1\lam_2\lam_3\lam_4)
\label{fields}\eeqa

\vskip 0.2in

  Let us  now start to check whether  or not all order higher derivative corrections of the D-brane Anti 
  D-brane system (that we derived in \reef{Lnm}) are exact. To achieve this aim, we eliminate all the over all factors from both string and field theory amplitudes and then test them at each order of $\alpha'$.
 At  $\alpha'$ order we get the following coefficient in field theory side :
\beqa
 (a_{1,0}-a_{0,1})(s'-t')-i(b_{1,0}-b_{0,1})(s'-t')&=&\pi^2(s'-t') \nonumber\eeqa

note that $b_{n,m}$ coefficients  are symmetric  and we used the coefficients in \reef{coefs}, which is exactly the same as $\pi^2  ( e_{1,0}-e_{0,1})(s'-t')$ that has been appeared in string amplitude. At  second order in alpha-prime or at  $\alpha'^2$ order , we get
\beqa
&& (a_{1,1}-ib_{1,1})(s't'-s't')-i(b_{2,0}-b_{0,2})(s'^2-t'^2)+(a_{2,0}-a_{0,2})[s'^2-t'^2]\nonumber\\
&&=4\pi^2 ln(2)(s'^2-t'^2)
\nonumber\eeqa
that is again the same as $\pi^2  ( e_{2,0}-e_{0,2})(s'^2-t'^2)$ in string amplitude, hence we can proceed further and notify the fact that all order $\alpha'$ corrections that we derived in \reef{Lnm} are exact. Thus not only were we able to produce all order alpha-prime corrections of two tachyon to fermion couplings in the world volume of D-brane Anti D-brane, but also we have shown that all the infinite tachyon singularities of 
$<V_C V_{\bar\Psi_1} V_{\Psi_2} V_{T_3} V_{T_4}>$
can be reconstructed as well. In \cite{Hatefi:2012cp} we have shown that the corrections of D-brane Anti D-brane system do differ from the corrections
of the non-BPS branes where for further details we invite the interested reader in following   \cite{Hatefi:2012cp}.

\vskip 0.2in

Notice to the important fact, if we make use of the coefficients  $a_{0,0}=2\ln(2)$ and $b_{0,0}=\pi/2$ then we immediately reveal that one is not able to match the non-Abelian couplings at $\alpha'$ order by taking  the symmetric trace of non-Abelian couplings of \reef{Lnm}.

\vskip 0.2in
 
Note that there is also an addition factor in DBI action which is a symmetrised trace on the $\sigma$ factors inside the so called DBI action, performing the symmetrised trace, one explores
\beqa
\frac{1}{2}\STr\left(V({ T^iT^i})\sqrt{1+[T^i,T^j][T^j,T^i]}\right)&=&\left(1-\frac{\pi}{2}T^2+\frac{\pi^2}{24}T^4+\cdots\right)\left(1+T^4+\cdots\right)\nonumber\eeqa
where the tachyon is going to condensate at $T\rightarrow \infty$  whereby the tachyon potential tends to zero at $T\rightarrow \infty$ indeed.


\vskip 0.2in

We also refer to some important points that are raised in favour of  the supersymmetric  action.  Let us end this section by just pointing out  the very crucial fact which is as follows.  Despite carrying out various researches on this subject, it is worth mentioning the fact that the entire form of symmetrised action is not found out yet \cite{Cederwall:1996pv,Cederwall:1996ri,Bergshoeff:1996tu}. 

 \section{$t-$ channel singularities }
 
Note that  regarding all the EFT methods and also considering the symmetries of the amplitude \cite{Schwarz:2013wra}, more crucially relying on  selection rules \cite{Hatefi:2013yxa}, we come to understand that there is no interaction between two fermions and a tachyon nor two tachyons and scalar field, therefore we have no $s',v',u',r'$ channel poles at all. The gauge field propagator is Abelian, but  the vertex of one RR, two fermion fields and a gauge field should be taken from the mixed RR and non-Abelian kinetic term of fermion fields. Hence,  one comes to know that there is no gauge field $w$ pole channel either.\footnote{  Although there is a non zero coupling between two tachyon and a gauge field that comes from kinetic term of tachyons in  DBI action.} One finds out the massless poles in an EFT as follows. \footnote{ The needed expansion for this part of the amplitude is given by  \beqa
-\frac{\pi^{3/2}}{t}-\frac{\pi^{7/2}}{6t}((s+u+v+r+1)^2+...)
\nonumber\eeqa.} All t-channel gauge field poles are generated for $p=n+2$ case by taking the following sub amplitude in an EFT
\beqa
V^{a,\alpha}(\bar \Psi_1,\Psi_2, A)G_{ab}^{\alpha\beta}(A)V^{b,\beta}(C_{p-3},A,T_3,T_4)\label{121}
\eeqa

 such that
 \beqa
  G_{ab}^{\alpha\beta}(A)&=&\frac{i\delta^{ab}\delta^{\alpha\beta}}{(2\pi\alpha')^2T_p t}\nonumber\\
 V^{a,\alpha}(\bar\Psi_1,\Psi_2, A)&=&T_p(2\pi\alpha')\bar u^A\ga^a_{AB}u^B \Tr(\lam_1\lam_2\lam^\alpha)\nonumber\\
V^{b,\beta}(C_{p-3},A,T_3,T_4)&=&i\mu'_p\beta (2\pi\alpha')^3\frac{1}{(p-2)!}  \sum_{n=-1}^{\infty}d_n \bigg(\alpha'k.(k_3+k_4)\bigg)^{n+1}\nonumber\\&&\times
  H_{a_0...a_{p-3}}k_{3a_{p-2}} k_{4a_{p-1}} \epsilon^{a_0...a_{p-1}b} \Tr(\lambda_3\lambda_4\lambda_{\beta})\label{kkio}
   \label{massless tchannel}
 \eeqa


 where the following Chern-Simons coupling has been taken into account

 \beqa
  && -\beta\mu'_p (2\pi\alpha')^{3}\int_{\Sigma_{p+1}} d^{p+1}\sigma \quad\Tr \bigg( C_{p-3}\wedge F\wedge DT\wedge DT \bigg) \label{mm1}\eeqa  

where the proper higher derivative corrections  are also applied to the above effective actions as follows
   
  \beqa
  && -\beta\mu'_p (2\pi\alpha')^{3}\int_{\Sigma_{p+1}} d^{p+1}\sigma   \sum_{n=-1}^{\infty} d_{n} \Tr \bigg( C_{p-3} \wedge D^{d_1}\cdots D^{d_{n}}   F \wedge  D_{d_1}\cdots D_{d_{n}}
  ( DT\wedge DT) \bigg)\nonumber\eeqa

 Replacing  \reef{massless tchannel} inside \reef{121} one finds out all t-channel gauge field singularities of the string amplitude as follows:
 
\beqa
 &&  \bar u^A\ga^a_{AB}u^B \frac{16\pi^2\mu'_p\beta}{(p-2)!}  \sum_{n=-1}^{\infty} \frac{1}{t} d_n \bigg(\alpha'k.(k_3+k_4)\bigg)^{n+1}\nonumber\\&&\times
 H_{a_0...a_{p-3}}k_{3a_{p-2}} k_{4a_{p-1}} \epsilon^{a_0...a_{p-1}a} \Tr(\lambda_1\lambda_2\lambda_3\lambda_4)\label{66}
   \label{bb4}
 \eeqa
 
     $k$ is the momentum of off-shell gauge field. Note that to obtain all those singularities we have used the fact that  $V^{a,\alpha}(\bar \Psi_1,\Psi_2, A)$ is derived from the fixed kinetic term of fermion fields in DBI action 
     as it does not receive any correction.  Therefore we applied the higher derivative corrections to Chern-Simons couplings to be able to exactly explore all  massless t-channel singularities of the amplitude for $p=n+2$ case in an EFT counterpart. 
      
\section{Conclusion }

 Let us conclude by arguing some general thoughts.   We have derived all order alpha-prime corrections of two fermion-two tachyon couplings in the world volume of D-brane Anti D-brane system. Not only were we able to produce these  corrections in \reef{Lnm} but also we have shown that all the infinite tachyon singularities of 
$<V_C V_{\bar\Psi_1} V_{\Psi_2} V_{T_3} V_{T_4}>$ can be reconstructed as well. We also  confirmed the presence of various new couplings  in the world volume of D-brane Anti D-brane system, which have new structures such as

\beqa
\bigg(\Psi^{(1)} D_aT\bar\Psi^{(2)} \ga^a  T^*\bigg), \quad \quad 
\bigg(\Psi^{(1)} T\bar\Psi^{(2)} \ga^a  D_aT^*\bigg)\label{88}\eeqa

 and in particular we explored their exact coefficients in \reef{Lnm}.
\vskip.2in

 We also derived  all  these  $\alpha'$ higher derivative corrections in type II in the presence of the constraint $p_ap^a\rightarrow 0$. Hence these couplings that we have gained, are applicable in the presence of this constraint, so these couplings cannot be compared with BSFT couplings. Nevertheless, tachyon's potential is used in superstring amplitudes as it  worked in BSFT, that is  $V({T})=e^{\pi\alpha'm^2{ T}^2}$ ~\cite{Kutasov:2000aq}, which can be represented by  \beqa
V( T^iT^i)&=&1+\pi\alpha'm^2{ T^iT^i}+
\frac{1}{2}(\pi\alpha'm^2{ T^iT^i})^2+\cdots
\non\eeqa  where 
$m^2=-1/(2\alpha')$ demonstrated  tachyon's mass, and tachyon condensation  was normally done at $T\rightarrow \infty$, whereby tachyon's potential is sent out to zero. Hence, 
once tachyon approaches to infinity
its potential will also be sent to zero, as one expected this result from tachyon condensation of a single brane.
\vskip.1in

These new corrections did play the important role  not only in revealing the structures of the singularities, but also in fixing the coefficients of higher derivative corrections of type II string theory.
It is  recently shown that how one can find corrections for Veneziano amplitude \cite{Veneziano:1968yb} for which the all order corrections to four tachyons are revealed in \cite{Hatefi:2017ags}. 

\vskip.1in

Finally, we have also fixed the  structures of fermionic corrections and also found couplings to all orders in the world volume of D-brane-Anti-D-brane system. Note that localization method was also used to deal with effective actions  \cite{Hashimoto:2015iha} and various fermionic interactions in the presence of Anti D3-branes for different fluxes have been discovered as well  \cite{Dasgupta:2016prs}.

 \section*{Acknowledgements}

Parts of the calculations of this paper were done at Mathematical Institute at Charles university, IHES in both physics and math departments, also at ICTP as well as at CERN where the author would like to show his gratitude to them for their warm hospitality. I would also like to thank  A. Sagnotti, N. Arkani-Hamed, B. Jurco, A. Sen, G.Veneziano, W. Siegel, P. Sundell, P. Horava, K. Narain, R. Russo, M. Douglas, D. Friedan, E. Martinec, S. Shenker as well as R. Antonelli and I. Basile for many useful discussions. I was supported in part by Scuola Normale Superiore and by INFN (ISCSN4-GSS-PI).

 \end{document}